\documentclass[twocolumn,showpacs,preprintnumbers,amsmath,amssymb,superscriptaddress]{revtex4}

\usepackage{graphicx}
\begin{document}
\bibliographystyle{prsty}
\title{Systematic changes of the electronic structure of the diluted ferromagnetic oxide Li-doped Ni$_{1-x}$Fe$_x$O with hole doping}

\author{M. Kobayashi}
\affiliation{Department of Physics, University of Tokyo, 
7-3-1 Hongo, Bunkyo-ku, Tokyo 113-0033, Japan}
\author{J. I. Hwang}
\affiliation{Department of Physics, University of Tokyo, 
7-3-1 Hongo, Bunkyo-ku, Tokyo 113-0033, Japan}
\author{G. S. Song}
\affiliation{Department of Physics, University of Tokyo, 
7-3-1 Hongo, Bunkyo-ku, Tokyo 113-0033, Japan}
\author{Y. Ooki}
\affiliation{Department of Physics, University of Tokyo, 
7-3-1 Hongo, Bunkyo-ku, Tokyo 113-0033, Japan}
\author{M. Takizawa}
\affiliation{Department of Physics, University of Tokyo, 
7-3-1 Hongo, Bunkyo-ku, Tokyo 113-0033, Japan}
\author{A. Fujimori}
\affiliation{Department of Physics, University of Tokyo, 
7-3-1 Hongo, Bunkyo-ku, Tokyo 113-0033, Japan}
\author{Y. Takeda}
\affiliation{Synchrotron Radiation Research Unit, 
Japan Atomic Energy Agency, Sayo-gun, Hyogo 679-5148, Japan}
\author{S.-I. Fujimori}
\affiliation{Synchrotron Radiation Research Unit, 
Japan Atomic Energy Agency, Sayo-gun, Hyogo 679-5148, Japan}
\author{K. Terai}
\affiliation{Synchrotron Radiation Research Unit, 
Japan Atomic Energy Agency, Sayo-gun, Hyogo 679-5148, Japan}
\author{T. Okane}
\affiliation{Synchrotron Radiation Research Unit, 
Japan Atomic Energy Agency, Sayo-gun, Hyogo 679-5148, Japan}
\author{Y. Saitoh}
\affiliation{Synchrotron Radiation Research Unit, 
Japan Atomic Energy Agency, Sayo-gun, Hyogo 679-5148, Japan}
\author{H. Yamagami}
\affiliation{Synchrotron Radiation Research Unit, 
Japan Atomic Energy Agency, Sayo-gun, Hyogo 679-5148, Japan}
\author{Y.-H. Lin}
\affiliation{State Key Laboratory of New Ceramics and Fine Processing, Department of Materials 
Science and Engineering, Tsinghua University, Beijing 100084, People's Republic of China}
\author{C.-W. Nan}
\affiliation{State Key Laboratory of New Ceramics and Fine Processing, Department of Materials 
Science and Engineering, Tsinghua University, Beijing 100084, People's Republic of China}\date{\today}

\begin{abstract}
The electronic structure of Li-doped Ni$_{1-x}$Fe$_x$O has been investigated using photoemission spectroscopy (PES) and x-ray absorption spectroscopy (XAS). The Ni $2p$ core-level PES and XAS spectra were not changed by Li doping. In contrast, 
the Fe$^{3+}$ intensity increased with Li doping relative to the Fe$^{2+}$ intensity. 
However, the increase of Fe$^{3+}$ is only $\sim 5\%$ of the doped Li content, suggesting that most of the doped holes enter the O $2p$ and/or the charge-transferred configuration Ni $3d^8\underline{L}$. 
The Fe $3d$ partial density of states and the host valence-band emission near valence-band maximum increased with Li content, consistent with the increase of electrical conductivity. 
Based on these findings, percolation of bound magnetic polarons is proposed as an origin of the ferromagnetic behavior.

\end{abstract}

\pacs{75.30.Hx, 75.50.Pp, 78.70.Dm, 79.60.-i}

\maketitle
In diluted magnetic semiconductors (DMS's), in which magnetic transition-metal ions are doped into non-magnetic semiconductor hosts, it is considered that the transition-metal ions magnetically interact with each other through itinerant carriers in the semiconductors. 
Carrier-induced ferromagnetism in diluted magnetic system is one of the most important properties for ``spin electronics" or ``spintronics" because this property enables us to control both the charge and spin degrees of freedom of electrons \cite{Science_98_Ohno, RMP_06_Jungwirth}.
Since the Curie temperature ($T_\mathrm{C}$) of typical ferromagnetic DMS Ga$_{1-x}$Mn$_x$As is lower than room temperature, it has been strongly desired to synthesize ferromagnetic DMS's having $T_\mathrm{C}$ above room temperature in order to utilize DMS's in practical applications. 
Ever since the reports on theoretical material design for high $T_\mathrm{C}$ DMS's \cite{Science_00_Dietl, JPCS_03_Yoshida}, the discovery of room-temperature ferromagnetism in DMS's based on wide-band gap or oxide semiconductors has been reported \cite{SST_04_Pearton, JMSME_05_Liu} although the origin of the ferromagnetism has not been clarified yet. 
Meanwhile, various attempts of other approaches to higher $T_\mathrm{C}$ have been made: These include nanostructure designing such as digital ferromagnetic heterostructures \cite{PRL_05_Nazmul, SSC_04_jeon} and nanoparticles \cite{PRL_03_Radovanovic, PRB_07_Karmakar}; employment of new host materials such as, e.g., layered compound Sb$_2$Te$_3$ \cite{PRB_06_Zhou}, correlated metal (La,Sr)TiO$_{3}$ \cite{PRL_06_Herranz}, and rare-earth oxide CeO$_{2}$ \cite{PRB_07_Fernandes} instead of conventional semiconductors.

Recently, Wang {\it et al.} \cite{APL_05_Wang} have taken another approach and reported ferromagnetic behaviors of Ni$_{1-x}$Fe$_x$O, where Fe atoms were doped into the antiferromagnetic insulator NiO instead of a non-magnetic semiconductor, with the $T_\mathrm{C}$ exceeding room temperature. 
A question then naturally arises what happens if such a system is doped with holes and whether the ferromagnetism is enhanced or not. 
Very recently, Lin {\it et al.} \cite{APL_06_Lin} have shown that the magnetization and electrical conductivity of Ni$_{1-x}$Fe$_x$O can be enhanced by Li co-doping. 
They have shown that the Li ions substituting the Ni sites generate holes and then act as acceptors. 
The magnetization increases with Li concentration while the electrical conductivity is nearly constant for finite Li concentrations, which implies that part of the doped holes may be compensated by defects such as oxygen vacancies. A core-level photoemission measurement revealed no metallic Fe signals and ruled out the possibility that the enhanced ferromagnetism is originated from the precipitation of Fe metal \cite{APL_06_Lin}. A previous x-ray absorption spectroscopy study on Fe$_{1-x}$Ni$_{x}$O \cite{JESRP_05_Chen} has demonstrated that even for high Ni concentrations ($x \gtrsim 0.7$), in which the structure is the rock-salt type like pure NiO, the Fe ions tend to be trivalent rather than divalent, probably due to the formation of cation vacancies. 
Because Li$_{y}$Ni$_{1-x-y}$Fe$_x$O (LNFO) is still semiconducting in spite of the increase of hole carrier concentration, the mechanism of the ferromagnetism is still unclear. 
In order to understand the origin of the ferromagnetism in LNFO, an understanding of the electronic structure including the valence of the Fe ion is required.

Photoemission spectroscopy (PES) and x-ray absorption spectroscopy (XAS) are powerful tools to investigate the electronic structure of materials. 
XAS, in which an electron is excited from a core level into unoccupied states, is an element specific technique and the spectra reflect the electronic structure associated with that element. 
Resonant photoemission spectroscopy (RPES) enables us to extract the $3d$ partial density of states (PDOS) of each $3d$ transition element. 
In this work, we have performed PES and XAS measurements on LNFO thin films having different hole concentrations controlled through Li doping in order to obtain fundamental information about the electronic structure. 
Based on the measured spectra, we shall discuss the origin of the ferromagnetism in LNFO.

\begin{figure}[b!]
\begin{center}
\includegraphics[width=8.0cm]{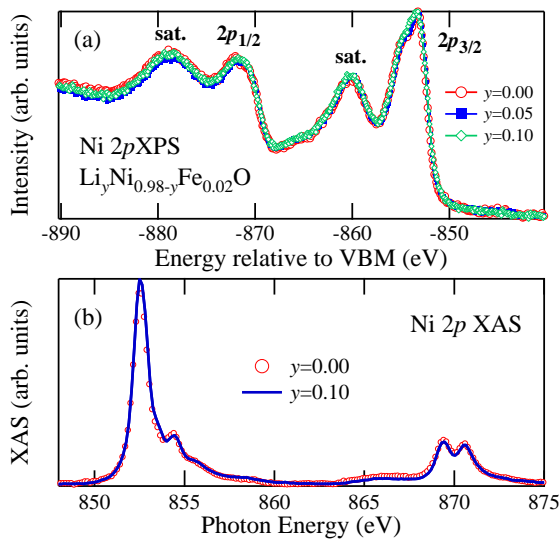}
\caption{Ni $2p$ core-level spectra of Li$_y$Ni$_{0.98-y}$Fe$_{0.02}$O. 
(a) Ni $2p$ core-level XPS spectra. 
(b) Ni $2p$ XAS spectra. 
}
\label{NiSpectra}
\end{center}
\end{figure}

Li$_y$Ni$_{0.98-y}$Fe$_{0.02}$O ($y$=0.00, 0.05, and 0.10) thin films were prepared on Si substrates by the sol-gel method combined with rapid thermal annealing. The total thicknesses of the films were $\sim50$ nm. X-ray diffraction and scanning electron microscopy were employed to reveal the microstructures and the phase compositions of the thin films. Ferromagnetism with $T_{\text{C}}$ above room temperature was confirmed by magnetization measurements using a SQUID magnetometer (Quantum Design, Co. Ltd.). Details of the sample fabrication and characterization were described in Ref.~\cite{APL_06_Lin}. 
RPES and XAS measurements were performed at the soft x-ray beam line BL23SU of SPring-8 \cite{AIP_04_Okamoto}. The monochromator resolution was $E/{\Delta}E \textgreater 10,000$. XAS signals were measured by the total electron yield method. 
X-ray photoemission spectroscopy (XPS) measurements were performed using a Gammadata Scienta SES-100 hemispherical analyzer and a Mg-$K\alpha$ x-ray source ($h{\nu}=1253.6$ eV). 
All the measurements were done at room temperature in the $10^{-8}$ Pa range. 
The total energy resolution of RPES and XPS measurements including temperature broadening was $\sim 300$ and $\sim 800$ meV, respectively. 
For surface cleaning, Ar$^+$ ion sputtering and annealing under oxygen pressure $\sim10^{-4}$ Pa were performed. Cleanness of the surface was checked by the absence of a high binding-energy shoulder in the O $1s$ core-level spectrum and the C $1s$ core-level contamination signal. 
Hereafter, photoelectron energies are referenced to the valence band maximum (VBM).

\begin{figure}[b!]
\begin{center}
\includegraphics[width=8.8cm]{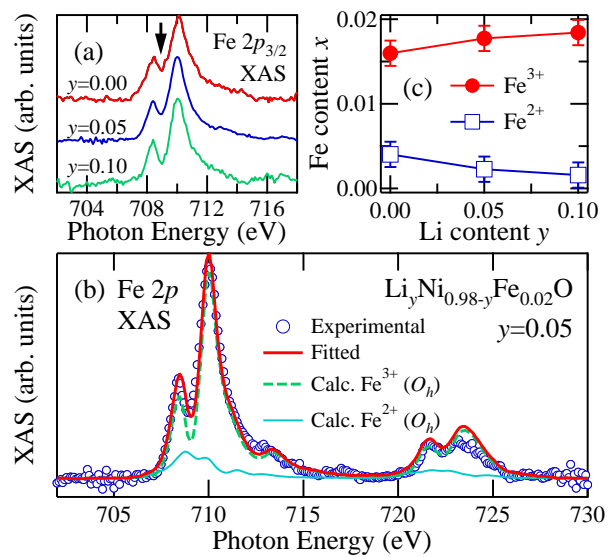}
\caption{Fe $2p$ XAS spectra of Li$_y$Ni$_{0.98-y}$Fe$_{0.02}$O. 
(a) Li concentration dependence. An allow shows the dip in the $2p_{3/2}$ region. 
(b) Fitting result for the 5\% Li sample, where the spectrum is assumed to be a superposition of the calculated spectrum of Fe$^{3+}$ and that of Fe$^{2+}$ in an $O_h$ crystal field \cite{PRB_95_Crocombette}. 
(c) Li concentration dependence of the Fe$^{3+}$ and Fe$^{2+}$ intensities plotted as functions of Li content. 
}
\label{FeL23XAS}
\end{center}
\end{figure}

Figure \ref{NiSpectra} shows the Ni $2p$ core-level XPS and XAS spectra of Li$_y$Ni$_{1-x-y}$Fe$_{x}$O ($x$=0.02). 
As shown in panel (a), the line shape of the XPS spectra of Li$_y$Ni$_{0.98-y}$Fe$_{0.02}$O is independent of Li content and similar to that of NiO. 
The Ni $2p$ XAS spectra of the LNFO samples are not changed by Li doping, as shown in panel (b), and were almost the same as that of pure NiO \cite{PRB_86_vanderLaan}. 
According to the previous reports on Li$_y$Ni$_{1-y}$O \cite{PRL_89_Kuiper, SSC_91_vanElp, PRB_92_vanElp, PRB_94_vanVeenendaal}, doped holes enter the ligand O $2p$ orbitals and the Ni$^{2+}$ ($d^8$) ground state turns into $d^8\underline{L}$ state, where $\underline{L}$ denotes a ligand hole, since NiO is a charge-transfer (CT) insulator \cite{PRB_84_Fujimori, PRB_86_vanderLaan}. 
Because the holes in the Ni $3d^8\underline{L}$ state reside in the ligand orbitals, the change of the Ni $2p$ XPS and XAS spectra with Li doping is small \cite{SSC_91_vanElp, PRB_92_vanElp, PRB_94_vanVeenendaal}. 
The observation that the Ni $2p$ XPS and XAS spectra of the Fe-doped samples are nearly independent of Li concentration is consistent with results on Li$_{y}$Ni$_{1-y}$O. 
The present result implies that the increase of electrical conductivity in LNFO is caused by holes having O $2p$ character.

\begin{figure}[t!]
\begin{center}
\includegraphics[width=8.8cm]{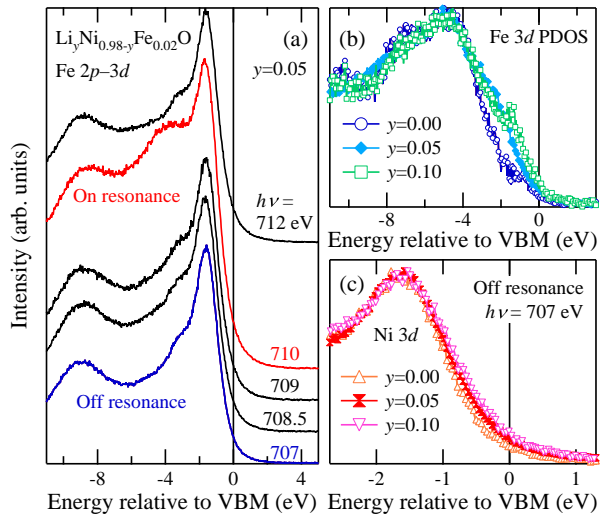}
\caption{Valence-band PES spectra of Li$_y$Ni$_{0.98-y}$Fe$_{0.02}$O. 
(a) Fe $2p \to 3d$ resonant photoemission spectra. The on- and off-resonance spectra were taken with $h\nu = 710$ and 707 eV, respectively. The difference between the on- and off-resonance spectra is the Fe $3d$ partial density of states (PDOS). 
(b) Fe $3d$ PDOS in Li$_y$Ni$_{0.98-y}$Fe$_{0.02}$O. 
(c) Off-resonance spectra of Li$_y$Ni$_{0.98-y}$Fe$_{0.02}$O. 
}
\label{Fe2p3dRPES}
\end{center}
\end{figure}

Li doping, on the other hand, affects the Fe $2p$ XAS spectra of the LNFO samples appreciably as shown in Fig.~\ref{FeL23XAS}. The two-peak structure in the Fe $2p_{3/2}$ XAS region ($h\nu=705-715$ eV) is characteristic of the Fe$^{3+}$ electronic configuration under an octahedral ($O_h$) crystal field. One can see from panel (b) that the dip in the $2p_{3/2}$ region becomes deeper with increasing Li concentration. 
Because the Fe $2p_{3/2}$ XAS spectrum of Fe$^{2+}$ ions should have a peak near the dip of the two-peak structure of Fe$^{3+}$, we consider that the Fe ions in Ni$_{1-x}$Fe$_x$O are mixture of predominant trivalent (Fe$^{3+}$) and a small amount of divalent (Fe$^{2+}$) states, and that Li doping converts part of Fe$^{2+}$ into Fe$^{3+}$. 
In a Mott insulator, a hole doped into an $N$-electron state generates different types of ($N-1$)-electron ground states, i.e., $d^n\underline{L}$ configuration for a CT insulator or $d^{n-1}$ configuration for a Mott-Hubbard (MH) insulator. 
The reason why the Fe$^{3+}$ ions are predominant already in the $y$=0.0 sample is probably because high concentration cation vacancies are created in Ni$_{1-x}$Fe$_{x}$O. 
Thus, the spectra have been fitted to a superposition of the calculated spectrum of Fe$^{3+}$ and that of Fe$^{2+}$ \cite{PRB_95_Crocombette}. 
The fitted spectra reproduce the XAS spectra well as shown in panel (b), and the concentration of the Fe$^{3+}$ and Fe$^{2+}$ components are obtained as shown in panel (c). 
Although the Fe$^{3+}$ concentration increased and the Fe$^{2+}$ one decreased with Li content, as expected, the amount of the increase of Fe$^{3+}$ is only $\sim 5\%$ of what would be expected for the doped Li content. 
This result means that a small fraction of doped holes are trapped by the Fe ions to form Fe$^{3+}$ while majority of uncompensated holes enter the O $2p$ band or the Ni $3d^8\underline{L}$ states.

In order to investigate corresponding changes induced by Li doping in the valence-band electronic structure, the doping dependence of the valence-band spectra are shown in Fig.~\ref{Fe2p3dRPES}. 
The Fe $3d$ PDOS has been extracted using the resonant photoemission technique. The spectra of Li$_y$Ni$_{0.98-y}$Fe$_{0.02}$O for photon energies in the Fe $2p \to 3d$ core-excitation region demonstrate a clear Fe $2p \to 3d$ resonance as shown in panel (a). 
Here, the on- and off-resonance photon energies have been chosen as $h\nu = 710$ and 707 eV, respectively, according to the Fe $2p$ XAS spectrum [Fig.~\ref{FeL23XAS}(a)]. The Fe $3d$ PDOS has been obtained from the difference between the on- and the off-resonance spectra. 
The Fe $3d$ PDOS shows a systematic change depending on the Li content as shown in panel (b), that is, the intensity near VBM increases with increasing Li concentration, where the PDOS spectra have been normalized to the peak height at $\sim -5$ eV relative to the VBM. Taking into account the increase of the Fe$^{3+}$ concentration with Li doping, the spectral change is likely related to the increases of the Fe$^{3+}$ concentration and the conductivity \cite{APL_06_Lin}. 
The off-resonance spectra, on the other hand, are almost the same as the valence-band PES spectra of NiO \cite{PRB_84_Fujimori, PRB_92_vanElp}. 
The off-resonance spectra near VBM representing the Ni $3d$ states (including Ni $3d^8\underline{L}$) demonstrate a small but systematic increase of intensity with Li content as shown in panel (c). The observations indicate that, apart from the Fe $3d$ orbital, Li doping affects the host valence band of NiO, too. 
It should be noted that with hole doping the density of states near VBM increases and that the conducting carriers have both Ni $3d^8\underline{L}$ and Fe$^{3+}$ character. This result is consistent with the picture of carrier-induced ferromagnetism in which the localized Fe $3d$ states and the itinerant NiO host band states are hybridized each other, resulting in certain Fe $3d$ character near VBM. 
Considering the fact that the electrical conductivity and the intensity of spectra near VBM increase with Li doping, a small amount of holes may go to the Fe $3d$ orbitals and be localized while the majority of holes go to the Ni $3d^{8}\underline{L}$ states hybridizing with the Fe $3d$ states.

Based on the present observations of the systematic changes of the electronic structure of LNFO with Li doping, we shall discuss possible mechanisms of the ferromagnetism. 
The magnetization of LNFO increases with Li content, and the electrical conductivity of LNFO is enhanced by almost four orders of magnitude compared with Ni$_{1-x}$Fe$_x$O \cite{APL_06_Lin}. 
Therefore, the ferromagnetism in LNFO is most likely due to a carrier-induced mechanism. 
Carrier-induced ferromagnetism in hole doped DMS may be categorized into two types: double exchange \cite{PR_51b_Zener, PRL_98_Akai} and $p$-$d$ exchange \cite{PR_51a_Zener, Science_00_Dietl}. 
In LNFO, since the doped holes are partially compensated and the LNFO thin films are semiconducting as described below, there are not sufficient itinerant carriers for the double exchange and $p$-$d$ exchange interactions to be very effective. 
In that case, relatively localized holes may form bound magnetic polarons \cite{PRL_02_Kaminski, NatMater_05_Coey} and their magnetic percolation. 
That is, exchange interaction between the magnetic ions and doped holes may lead to the formation of percolated bound magnetic polarons at low carrier density. 
In addition to the Fe ions, the doped Li ions and cation vacancies are candidates for the hole localization centers in LNFO. 
The increase of magnetization with Li concentration may be related to the number of bound magnetic polarons and overlap between them. 
If the exchange interaction between the local magnetic moments and the hole spins is stronger than the superexchange interaction between neighboring Ni ions, within the bound magnetic polaron the magnetic moments of the Ni ions will be parallel to that of the Fe ions. 
In Li-free Ni$_{1-x}$Fe$_{x}$O, the Fe atoms and cation vacancies may also become the origin of the hole carriers and the centers of hole localization. 
In order to obtain fundamental understanding of the ferromagnetism, magnetoresistance and x-ray magnetic circular dichroism measurements on LNFO are highly desired.

In LNFO, the changes of the electronic state of Fe and the increase of the intensity near VBM with Li doping can be explained by the MH nature of the Fe$^{2+}$ ions and the CT nature of the Ni$^{2+}$ ions. 
Therefore, the classification of doped transition-metal ions according to the Zaanen-Swatzky-Allen scheme \cite{PRL_85_Zaanen} is a useful basis because doped holes tend to be localized in the MH-type compounds and itinerant in the CT-type ones in hole-doped diluted magnetic systems. 
Indeed, it has been reported that the Mn ions in Ga$_{1-x}$Mn$_{x}$As and In$_{1-x}$Mn$_{x}$As can be classified into the CT type \cite{PRB_99_Okabayashi, PRB_02_Okabayashi}, resulting in hole carriers of $p$-type character, meanwhile the Cr ions in Ga$_{1-x}$Cr$_{x}$N can be classified into the MH type \cite{DrThesis_06_Hwang}, resulting in hole carriers of $d$-type character.

In conclusion, we have performed PES and XAS measurements on Li$_y$Ni$_{0.98-y}$Fe$_{0.02}$O thin films with various Li concentrations. 
While the Ni $2p$ XPS and $2p$ XAS spectra hardly depended on Li content, the Fe $2p$ XAS spectra and the Fe $3d$ PDOS in the valence band showed systematic changes with Li concentration, indicating that the hole doping affects the electronic structure of the Fe ions. The Fe$^{3+}$ intensity relative to the Fe$^{2+}$ one, however, increased only slightly with Li concentration, suggesting that most of the carriers enter the host NiO band. 
The Fe $3d$ PDOS and the host valence-band spectra near VBM were enhanced with Li concentration, supporting the idea of carrier-induced ferromagnetism in this system. 
The changes in the electronic structure around the Fe and Ni ions with Li doping can be explained within the Zaanen-Sawatzky-Allen diagram for Mott insulators. 
Based on the experimental findings, we suggest that the carrier-induced ferromagnetic properties of Li$_{y}$Ni$_{1-x-y}$Fe$_x$O are caused by the formation of bound magnetic polarons, consisting of doped holes in the Ni $3d^8\underline{L}$ state and the Fe local magnetic moments.

This work was supported by a Grant-in-Aid for Scientific Research in Priority Area ``Invention of Anomalous Quantum Materials'' (16076208) from MEXT, Japan. YHL and CWN acknowledge support from NSF of China (50502024 and 50621201). MK and MT acknowledge support from the Japan Society for the Promotion of Science for Young Scientists.

\end{document}